# Nanoscale strain manipulation of smectic susceptibility in kagome superconductors

Yidi Wang[1, *], Hong Li[1, *], Siyu Cheng[1], He Zhao[1], Brenden R. Ortiz[2], Andrea Capa Salinas[2], Stephen D. Wilson[2], Ziqiang Wang[1] and Ilija Zeljkovic[1]

[1] *Department of Physics, Boston College, Chestnut Hill, MA 02467, USA*

[2] *Materials Department, University of California Santa Barbara, Santa Barbara, California 93106, USA*

*\* Equal contribution*

*Correspondence:* ilija.zeljkovic@bc.edu

**Exotic quantum solids can host electronic states that spontaneously break rotational symmetry of the electronic structure, such as electronic nematic phases and unidirectional charge density waves (CDWs). When electrons couple to the lattice, uniaxial strain can be used to anchor and control this electronic directionality. Here we reveal an unusual impact of strain on unidirectional "smectic" CDW orders in kagome superconductors $AV_3Sb_5$ using spectroscopic-imaging scanning tunneling microscopy. We discover local decoupling between the smectic electronic director axis and the direction of anisotropic strain. While the two are generally aligned along the same direction in regions of small CDW gap, the two become misaligned in regions where CDW gap is the largest. This in turn suggests nanoscale variations in smectic susceptibility, which we attribute to a combination of local strain and electron correlation strength. Overall, we observe an unusually high decoupling rate between the smectic electronic director of the 3-state Potts order and anisotropic strain, revealing weak smecto-elastic coupling in the CDW phase of kagome superconductors. This is phenomenologically different from the extensively studied nemato-elastic coupling in the Ising nematic phase of Fe-based superconductors, providing a contrasting picture of how strain can control electronic unidirectionality in different families of quantum materials.**

## Introduction

In quantum materials with substantial correlation effects, electrons often form phases with reduced symmetry groups compared to that of the underlying lattice structure. These for example include electronic nematic phases in high-temperature superconductors and Moire systems[1–3], where certain in-plane mirror symmetries are broken, resulting in an electronic band structure that is different along nominally equivalent lattice directions. Another canonical example is the stripe-like charge ordering in cuprate high-temperature superconductors, where both rotational and translational symmetry are simultaneously broken to form a "smectic" electronic phase[1].

The family of $A$V$_3$Sb$_5$ ($A$ stands for K, Cs or Rb) kagome superconductors recently emerged as rich playground to study symmetry-broken electronic phenomena[4–31]. In particular, the $2a_0 \times 2a_0$ charge density wave (CDW) phase (onset temperature $T^* \sim$ 80-100 K[32–34]) that appears to break time-reversal symmetry[4,7,18,23] attracted tremendous attention for potential realization of exotic orbital loop currents[12,35–37]. This phase also breaks the six-fold rotational symmetry of the lattice[4,6,7,25,38], with a single in-plane reflection symmetry that remains, suggesting a unidirectional smectic nature of the phase.

In materials where electrons strongly couple to the lattice, uniaxial strain can be a powerful knob for controlling the directionality of symmetry breaking phenomena[39–46], by for instance anchoring the nematic axis[39–41,47,48]. These experiments typically rely on either applying a hydrostatic pressure that simultaneously compresses all crystalline directions, or exerting in-plane strain which in turn allows the out-of-plane lattice parameter to respond such that volume is approximately conserved. The majority of explorations have been focused on solids with square or tetragonal in-plane lattice structures such as Fe-based superconductors, characterized by Ising-like nematic order parameter. Unlike these systems, $A$V$_3$Sb$_5$ has a hexagonal in-plane lattice structure, in which rotational symmetry breaking states may be characterized by a 3-state Potts model[10,46,49]. Previous pressure and strain studies utilized similar non-local bulk techniques, predominantly focusing on studying the $2a_0 \times 2a_0$ CDW phase, and the competition between CDWs and superconductivity[21,50–62]. However, there still exists little information related to if and how one can use strain to control smectic CDW domains, and if it is possible to rotate the director axis associated with the order. Understanding strain tunability, which reflects the underlying strength of electron-lattice coupling, holds important insights into the nature of this state.

Here we study the impact of strain on the smectic CDWs in kagome superconductors CsV$_3$Sb$_5$ and KV$_3$Sb$_5$. We focus on the occasionally observed buckled regions of the sample, and use a phase-sensitive analysis applied to atomically resolved STM topographs to determine local in-plane strain. Our experiments reveal that the $4a_0$ charge stripe order in CsV$_3$Sb$_5$ can be easily suppressed by strain. Moreover, we discover substantial deviations between the direction of the smectic electronic director of the $2a_0 \times 2a_0$ CDW and the direction of in-plane uniaxial strain, suggesting weak smecto-elastic coupling in kagome superconductors. By studying different strained regions in more detail, we find that the misalignment of the electronic smectic director and anisotropic in-plane strain generally occurs in the regions with large CDW gap. Our experiments provide an atomic-scale insight into the unusual interplay of strain and smectic orders in this family of materials, different from other commonly studied materials with a pronounced electronic directionality.

**Results**

Bulk single crystals of $A$V$_3$Sb$_5$ cleave between hexagonal Sb layers and $A$ layers (inset in Fig. 1a, Methods). We focus on the Sb surface, which has a hexagonal morphology with in-plane lattice constants of $a \approx b \approx 5.4$ Å consistent with diffraction measurements[34]. The V kagome layer, an essential ingredient in the crystal structure of these materials, resides just below the Sb surface imaged. Similarly to other experiments on CsV$_3$Sb$_5$[5,22,38,63,64], typical STM topographs of flat regions of the sample show both the $2a_0 \times 2a_0$ CDW modulations and the $4a_0$ charge-stripe order (Fig. 1b). We proceed to explore how these smectic orders can be controlled by strain.

For this purpose, we focus on the occasionally encountered strained regions of the sample. In Fig. 1a, we show an area with a single bright "ripple" running across the field-of-view. By closely examining the topographic signal over this region and away from it, we find that the $4a_0$ charge stripe order is markedly suppressed over the ripple (Fig. 1d-g). This is supported by the FT of both the flat and the rippled region, where the $\mathbf{q}_{4a0}$ peak appears in the former (Fig. 1b,c) and disappears in the latter (Fig. 1d,e). Such buckled regions of the sample are generally induced by accidental strain[42,47,65]. To investigate this quantitatively, we apply the Lawler-Fujita drift-correction algorithm[66] on the topograph to extract the spatially varying strain tensor[47,48,66–69]. From the $x$ and $y$ components of the displacement fields, $u_x(\mathbf{r})$ and $u_y(\mathbf{r})$, we can extract the strain tensor as:

$$\mathbf{u}(\mathbf{r}) = \begin{pmatrix} u_{xx}(\mathbf{r}) & u_{xy}(\mathbf{r}) \\ u_{yx}(\mathbf{r}) & u_{yy}(\mathbf{r}) \end{pmatrix} = \begin{pmatrix} \partial_x u_x(\mathbf{r}) & \frac{1}{2}(\partial_y u_x(\mathbf{r}) + \partial_x u_y(\mathbf{r})) \\ \frac{1}{2}(\partial_y u_x(\mathbf{r}) + \partial_x u_y(\mathbf{r})) & \partial_y u_y(\mathbf{r}) \end{pmatrix}$$

where $u_{xx}(\mathbf{r})$ and $u_{yy}(\mathbf{r})$ are the strain map components along the $x$ and $y$ axes, and $u_{xy}(\mathbf{r}) = u_{yx}(\mathbf{r})$ are the shear strain components. Through basis transformation of the strain tensor matrix, we can extract strain components along any direction (Supplementary Note 1). In particular, we can directly measure strain along $\mathbf{Q}_{\text{Bragg}}^a$, $\mathbf{Q}_{\text{Bragg}}^b$ and $\mathbf{Q}_{\text{Bragg}}^c$ lattice directions independently, denoted as $u_{aa}(\mathbf{r})$, $u_{bb}(\mathbf{r})$, and $u_{cc}(\mathbf{r})$ (Figure 1h-j). Positive values represent relative tensile strain while negative values represent relative compressive strain. We note that these values are relative to the area away from the ripple, which we assume to be characterized by approximately zero strain. The most pronounced strain variation is observed along the $\mathbf{Q}_{\text{Bragg}}^b$ direction, which is directly perpendicular to the ripple. We calculate cross-correlation coefficients between the $4a_0$ charge stripe order intensity map and different strain map components to find the highest positive correlation for the $\mathbf{Q}_{\text{Bragg}}^b$ direction (Fig. 1k, Supplementary Note 2). This suggests that strain parallel to the $4a_0$ wave vector, possibly in combination with out-of-plane strain, leads to the suppression of the order.

We now turn to the impact of strain on the parent $2a_0 \times 2a_0$ CDW phase. Previous STM experiments thoroughly investigated the anisotropy of the CDW state to find it is characterized by a dominant CDW direction different from the other two that are approximately the same[6,7,25,38,63]. This suggests that rotation symmetry of the CDW state is reduced from $C_6$ to $C_2$ with a single remaining in-plane reflection symmetry along the dominant direction – a "smectic" CDW state. We define a smectic axis order parameter associated with the $2a_0 \times 2a_0$ CDW, equivalent to the nematic vector order parameter for the 3-Potts nematic state[10], which can be extracted locally from the intensities of the three CDW peaks $I_{2a_0}^{a,b,c}$ as: $\mathbf{n} = (I_{2a_0}^a + I_{2a_0}^c - 2I_{2a_0}^b, \sqrt{3}(I_{2a_0}^c - I_{2a_0}^a))$. This yields three main vector directions $\mathbf{n}$: $(1,0), (1,\sqrt{3})$ and $(1,-\sqrt{3})$ corresponding to $\mathbf{q}_{2a_0}^b$, $\mathbf{q}_{2a_0}^c$ and $\mathbf{q}_{2a_0}^a$, respectively.

For simplicity, we first examine the simpler case of KV$_3$Sb$_5$, where the 4$a_0$ stripe order is generally absent (Fig. 2a,b). We analyze a region with strong spatial modulations and extract strain tensor components from the topograph to map local strain (Fig. 2c-e). Strain appears to show strongest variations along the $Q_{\text{Bragg}}^c$ direction, with the spatial distribution qualitatively similar to those seen in the topograph. We create the intensity map of each 2$a_0$ × 2$a_0$ CDW peak $q_{2a0}^i$ over the same region (Fig. 2f-h) and cross-correlate with the strain maps. The analysis shows a high positive cross-correlation coefficient between the strain along $Q_{\text{Bragg}}^c$ and the intensity of the $q_{2a0}^c$. In other words, stretching the material along one lattice direction leads to a concomitant change in the amplitude of the CDW peak along that same direction. Interestingly however, strain in this region does not alter the direction of the smectic director axis $n$, which points approximately along $Q_{\text{Bragg}}^b$ in the whole field of view despite in-plane strain of several percent stretching the lattice in the direction of $Q_{\text{Bragg}}^c$ (Fig. 2l).

To investigate this further, we proceed to examine an STM topograph of CsV$_3$Sb$_5$ where we find a more dramatic variation of electronic properties (Fig. 3a). The FT of the STM topograph shows both the $q_{2a0}$ and $q_{4a0}$ wave vectors (Fig. 3b). Similarly to what we discussed in Fig. 1, the 4$a_0$ stripe order disappears in the middle region due to strain (Fig. 3a, Supplementary Figure 1). Spatial variations of the smectic director axis can be determined by a visual inspection of a d$I$/d$V$($r$, $V$=-20 mV) map (Fig. 3c). We pinpoint several distinct regions, labeled in alphabetical order from A to E, outlined by dashed lines across which the smectic director axis rotates towards a different Bragg peak direction. The identification of the rotation axis for each domain is further supported by the CDW amplitude profiles (Supplementary Figure 2).

Another insight comes from measuring the spectral gap variations. The gap at the Fermi level measured in the average d$I$/d$V$ spectrum is about 35 meV, which agrees with the magnitude of the 2$a_0$ × 2$a_0$ CDW gap reported in previous STM work[4–6,22]. This local CDW gap, however, varies in different regions. A linecut of d$I$/d$V$ spectra acquired across individual regions A, B, C and D illustrates the gap size modulation (Fig. 3d). From this linecut, we can further exclude that the dominant spectral gap arises from the 4$a_0$ charge stripe order, as nearly identical gap edges are observed in regions A and B, with and without the 4$a_0$ stripe order. By mapping the gap size for each pixel in the field-of-view, we create a gapmap (Fig. 3e, Supplementary Note 3), which shows tantalizing similarities to the smectic domain morphology. Domain C contains the largest CDW gap, whereas domain A exhibits the smallest ones (Fig. 3f).

Similar to the procedure used in Figs. 1 and 2, we extract in-plane strain maps $u_{aa}(r)$, $u_{bb}(r)$, and $u_{cc}(r)$ from the STM topograph in Fig. 3a (Fig. 4a-c, Supplementary Figure 3). To visualize if the area is locally deformed isotropically or anisotropically, we plot a real-space lattice deformation map based on the full strain tensor $u(r)$ (Fig. 4i, Supplementary Note 4). In this map, isotropic disks represent either isotropic or zero strain, while elongated ellipses denote tensile strain that is parallel to the major axis of the ellipse. The relative variation of the area for each disk reflects the change of the in-plane unit cell area. In general, the more deformed the disk is, the more pronounced the corresponding in-plane strain is. To accompany local strain mapping, we plot the direction of the smectic director $n$ (Fig. 4g) and the associated angles $\theta(n)$ over the identical region of the sample to visualize smectic domains (Fig. 4h). In the $\theta(n)$ map, there are three different majority vectors at 29.6°±0.1°, 92.8°±1.8° and 151.0°±0.3°, which corresponds to the three possible directions of the smectic order parameter. This method provides a more quantitative measurement of smectic domains, which are generally consistent with the domain walls picked out by visual inspection of data in Fig. 3c.

Overlaying domain walls over the deformation map reveals important insights. Domain A away from the rippled area (upper left and lower right corner of Fig. 4i) exhibits the $4a_0$ charge-stripe order consistently seen on the nominally unstrained surfaces of CsV$_3$Sb$_5$, suggesting minimal strain. None of the other regions in the field-of-view show the $4a_0$ charge order. Interestingly, domain B shows two types of tensile strain: stretching along $\boldsymbol{Q}_{\text{Bragg}}^{b}$ (left part) and stretching along $\boldsymbol{Q}_{\text{Bragg}}^{c}$ (right part), but maintains the $\boldsymbol{q}_{2a0}^{b}$ smectic axis throughout. This suggests that tensile strain aligns with the direction of the smectic axis only in a part of this domain. Notably, domain C shows the electronic smectic axis along $\boldsymbol{q}_{2a0}^{c}$ despite tensile strain along $\boldsymbol{Q}_{\text{Bragg}}^{b}$. On the other hand, smectic director in domain D is the same as the direction of tensile strain. Overall, our data in Fig. 2, Fig. 4 and Supplementary Figure S4 (a) demonstrates that the $C_2$ smectic CDW director axis can rotate in response to uniaxial strain in only some of the nanoscale regions.

**Discussion**

To shed light on this intriguing observation, we turn to the variation of spectral gaps between different regions (Fig. 4j). In particular, we can observe that regions where $C_2$ electronic smectic director axis and uniaxial strain are misaligned generally show a larger CDW gap, for example domain C and the right half of domain B. In analogy to the definition of nematic susceptibility, we can define smectic susceptibility $\chi_{\text{smec}} = \frac{\partial \bar{n}}{\partial u}$ ($\bar{n}$ is the average smectic vector) in a 3-state Potts model. We can write the Hamiltonian for the system as $H = -\sum_{<i,j>} J_{ij} n_i n_j - \sum_i u_i n_i$, where $n_i$ is the site-dependent smectic vector, $J_{ij}$ is the site-dependent coupling constant and $u_i$ is the site-dependent magnitude of tensile strain. When $\chi_{\text{smec}}$ is large, it is energetically costly to misalign the smectic vector and the strain field; however, when $\chi_{\text{smec}}$ is small, the energy loss of the misalignment between local smectic vector and the strain field can be compensated by the gain in the coupling energy of neighboring smectic vectors themselves. From this we can see that the "hard" $C_2$ smectic electronic axis (with small $\chi_{\text{smec}}$) will be difficult to be rotated by strain along a different direction, analogous to a hard magnet that resists the saturation of the magnetization in magnetic field.

Our STM experiments shed light on the relationship between smectic CDWs and strain in kagome superconductors. We find that strain can induce a complete suppression of the $4a_0$ charge-stripe order in CsV$_3$Sb$_5$ while the $2a_0 \times 2a_0$ CDW remains present. The fact that strain leads to the disappearance of the $4a_0$ charge stripe order before the $2a_0 \times 2a_0$ CDW is also consistent with the smaller onset temperature of the $4a_0$ charge order compared to that of the $2a_0 \times 2a_0$ CDW[5]. Interestingly, the spontaneously formed $C_2$ smectic director axis of the $2a_0 \times 2a_0$ CDW does not generally follow the direction of uniaxial strain. In principle, the smectic vector could choose one of the 3 degenerate states most closely aligned with the direction of the tensile strain to minimize the total energy. We hypothesize that stronger electron correlation driving the larger CDW gap in turn substantially reduce smectic susceptibility so that in-plane uniaxial strain field fails to rotate the overall smectic vector. This variation of the CDW gap can in part be attributed to the modulations in the out-of-plane strain, i.e. interlayer distance, as revealed by bulk-sensitive strain measurements[21,62,70], where the CDW gap increases when interlayer distance increases.

It is interesting to compare our observations to the extensive elasto-transport experiments studying the coupling of electrons and the approximately square lattice in the nematic phase of Fe-based superconductors[40,45,48]. While two types of "twin" electronic nematic domains are routinely observed, these can be easily de-twinned by small amounts of uniaxial strain of less than 1% [40,71]. In contrast in $AV_3Sb_5$, it appears that substantially larger strain does not consistently align the smectic director axis, which could be an important insight for elasto-transport measurements on these systems. Locally strained regions in $AV_3Sb_5$ provide a fortuitous platform to study the effects of strain using a local probe to test the stiffness of smectic CDWs. This could serve as a useful methodology to provide fresh insights into the control of electronic directionality in other correlated materials.

**Methods**

Single crystals of $CsV_3Sb_5$ and $KV_3Sb_5$ were grown and characterized as described in more detail in Ref. [32,34]. We cold-cleaved the crystals at low-temperature as described in Ref. [5] and quickly inserted them into STM head. STM data was acquired using a customized Unisoku USM1300 microscope at approximately 4.5 K. Spectroscopic measurements were made using a standard lock-in technique with 910 Hz frequency and bias excitation as also detailed in figure captions. STM tips used were home-made chemically etched tungsten tips, annealed in UHV to bright orange color prior to STM experiments.

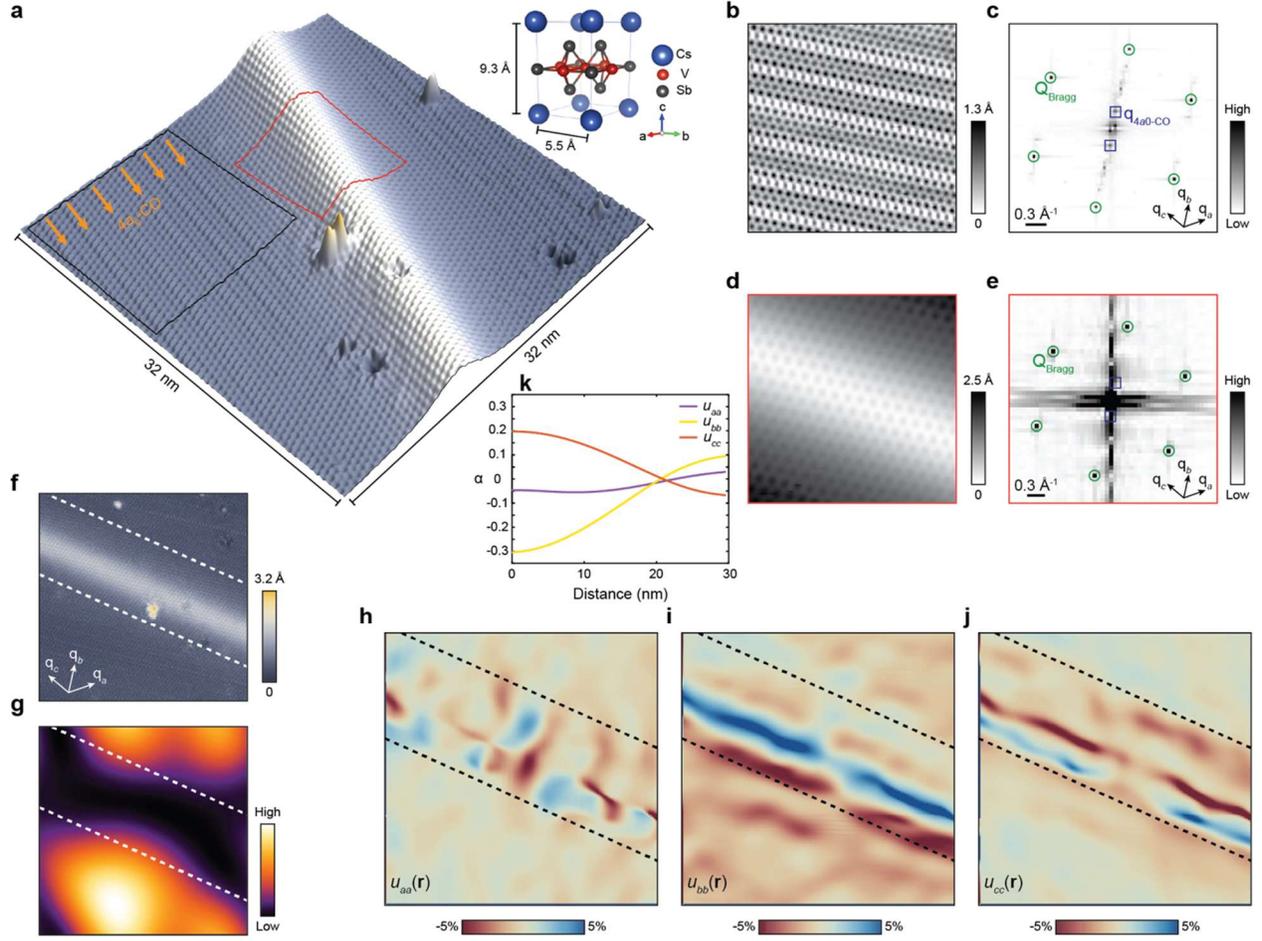

**Figure 1. The suppression of the $4a_0$ charge stripe order under strain. a,** A 3D portrayal of a large-scale topograph of a surface-corrugated region of the Sb termination. The orange arrows denote the $4a_0$ stripe order. The upper right inset shows the crystal structure. **b,** STM topograph of the black square region in (a) with **c,** its associated Fourier Transform (FT), where Bragg peaks are circled in green and $q_{4a_0}$ peak is enclosed in blue squares. **d,** STM topograph of the red square region in (a) and **e,** its FT, in which $q_{4a_0}$ peak disappears. We note that the $2a_0$ by $2a_0$ CDW peaks are difficult to be discerned in (e) but they are clearly visible in the Fourier transform of the entire region shown in (a). **f,** STM topograph of the same region as (a), with white dashed lines marking the region without $4a_0$ stripe order. **g,** $4a_0$ stripe order intensity map of (f). **h-j,** Strain maps of (f) along three reciprocal lattice directions $Q_{Bragg}^{a}$, $Q_{Bragg}^{b}$, and $Q_{Bragg}^{c}$, respectively. The black dashed lines denote the same non-$4a_0$ region in (f) and (g). Positive (negative) values indicate relative tensile (compressive) strain. **k,** Radially averaged cross-correlation coefficients α between the $4a_0$ intensity map (g) and strain maps (h-j), plotted as a function of distance. The cross-correlation coefficient between two images $A(x,y)$ and $B(x,y)$ is defined as $\alpha(u,v) = \frac{\sum_{x,y} A(x,y)B(x+u,y+v)}{\sum_{x,y} A^2(x,y) \sum_{x,y} B^2(x+u,y+v)}$ where $u$, $v$ are shifts along $x$ and $y$ directions, respectively. Purple, yellow, and red curves represent the correlation along $Q_{Bragg}^{a}$, $Q_{Bragg}^{b}$, and $Q_{Bragg}^{c}$, respectively. STM setup condition: $V_{sample} = 200$ mV, $I_{set} = 100$ pA.

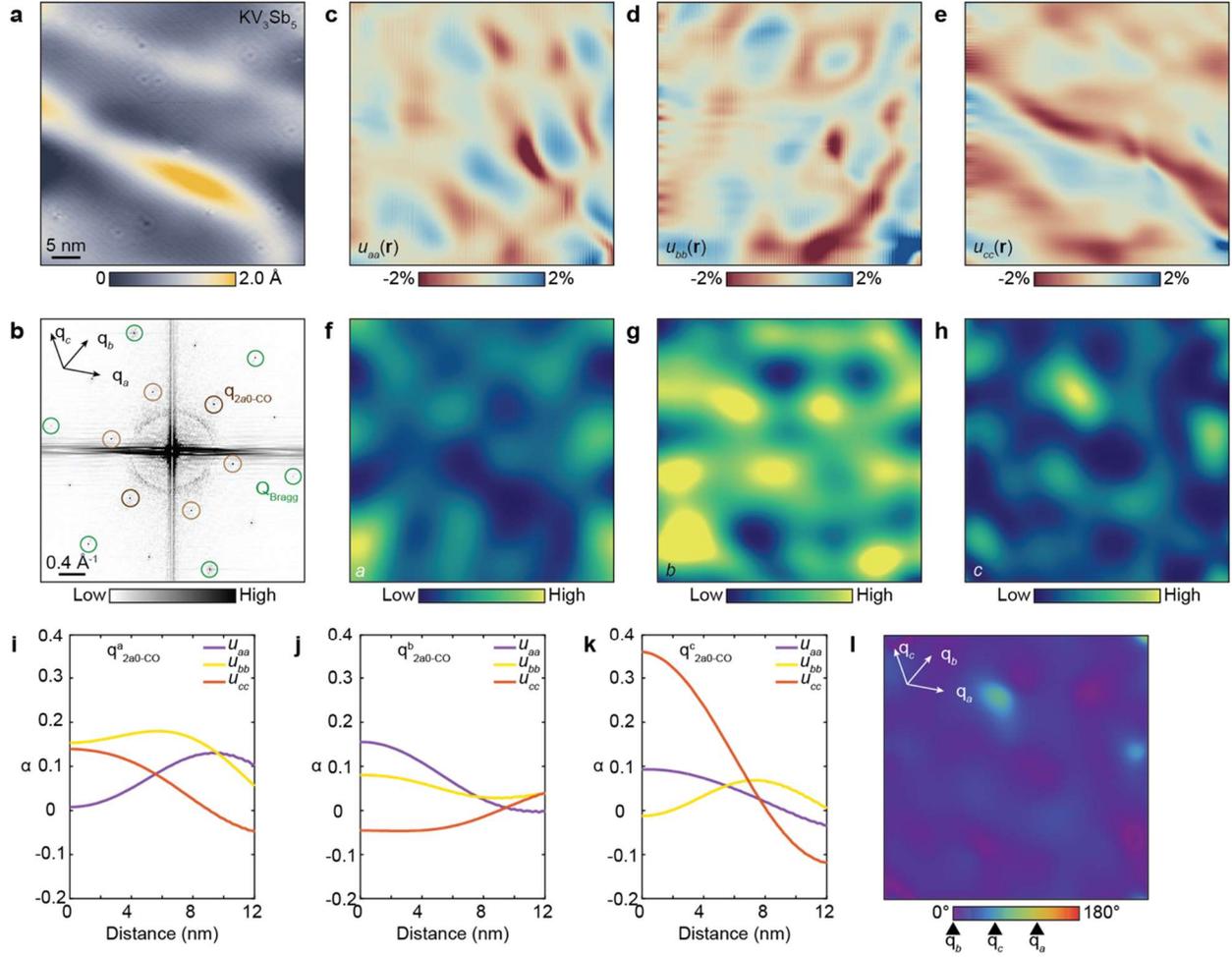

**Figure 2. Modulation of $2a_0$ CDW intensity under strain. a, b,** STM topograph of an Sb-terminated $KV_3Sb_5$ surface corrugated by strain and its FT. Green and brown circles indicate the atomic Bragg peaks and $\mathbf{q}_{2a0}$ peaks, respectively. **c-e,** Strain maps over (a) along three reciprocal lattice directions after linear background subtraction to remove tip drift. **f-h,** $2a_0$ CDW intensity maps along three reciprocal lattice directions, extracted from (a) and plotted in the same scale. **i-k,** Radially averaged cross-correlation coefficients between strain maps (c-e) and $2a_0$ CDW intensity maps (f-h), plotted as a function of distance. The color notation is the same as Figure 1(k). **l,** Angle map of the smectic vector order defined as $\mathbf{n}(\mathbf{r}) = (I_{2a_0}^a(\mathbf{r}) + I_{2a_0}^b(\mathbf{r}) - 2I_{2a_0}^c(\mathbf{r})$, $\sqrt{3}(I_{2a_0}^b(\mathbf{r}) - I_{2a_0}^a(\mathbf{r}))$, where $I_{2a_0}^a(\mathbf{r})$, $I_{2a_0}^b(\mathbf{r})$ and $I_{2a_0}^c(\mathbf{r})$ stand for CDW intensity along $\mathbf{q}_{2a0}^a$, $\mathbf{q}_{2a0}^b$ and $\mathbf{q}_{2a0}^c$, respectively. Angle of each smectic vector $\mathbf{n}(\mathbf{r})$ is measured with respect to $\mathbf{q}_{2a0}^b$ direction, and black triangles under the color bar denote the color representations of each reciprocal lattice direction. STM setup condition: $V_{\text{sample}} = 100$ mV, $I_{\text{set}} = 200$ pA.

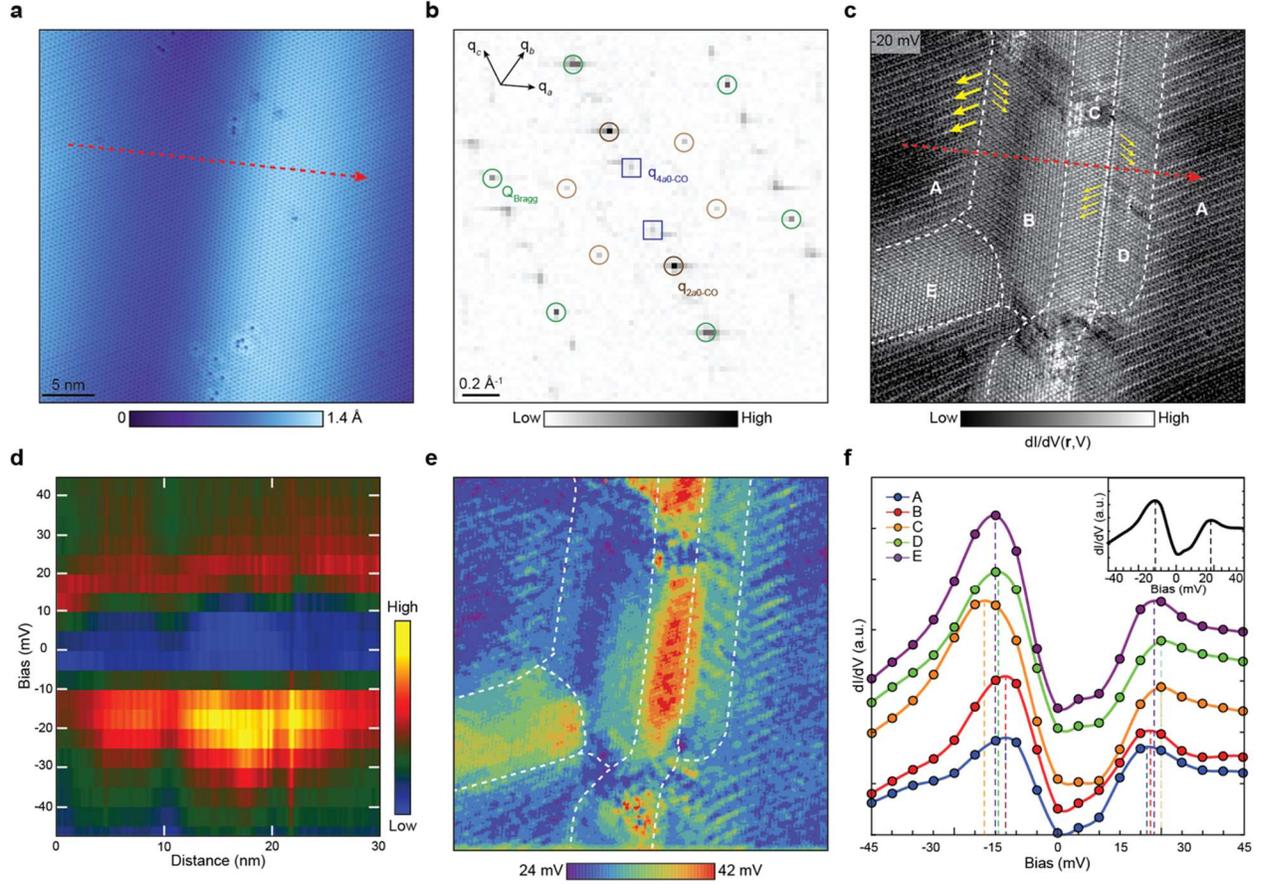

**Figure 3. Modulation of $2a_0$ CDW gap and rotation symmetry. a, b,** STM topograph of Sb-terminated surface of $CsV_3Sb_5$ and its FT. Green circles, brown circles and blue squares indicate the atomic Bragg peaks, $q_{2a0}$ and $q_{4a0}$ peaks, respectively. **c,** Representative $dI/dV(r, V=-20\ mV)$ map of the same region as (a). Five domains A-E are determined by the wave vector length and directions of the eye-caught stripe order. Domain wall is marked by white dashed lines. Yellow arrows denote the local stripe order directions. **d,** A 30 nm long $dI/dV$ spectra real space linecut along the red dashed arrow in (a) and (c). **e,** $2a_0$ CDW gap map over the same region as (a). The gap size at each point $r$ is determined by extracting the energy of local maximum for coherent peaks in $dI/dV$ spectra near the Fermi level. **f,** Spatially averaged $dI/dV$ spectra acquired over the five domains denoted by different colors. The dashed lines mark the peak positions, signifying the gap size. Five curves are offset for clarity. The upper right inset demonstrates the spatially averaged $dI/dV$ spectra over the same region as (a). STM setup condition: $V_{sample} = 100\ mV$, $V_{exc} = 5\ mV$, $I_{set} = 300\ pA$.

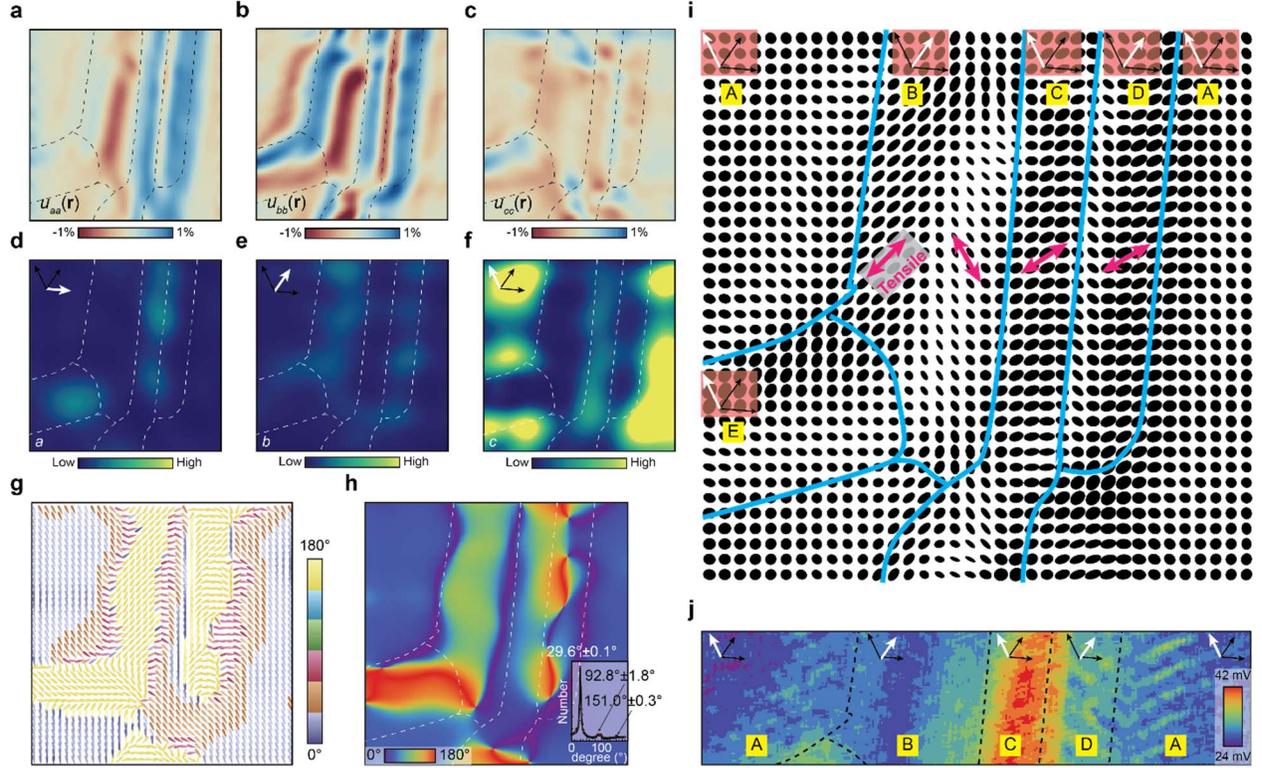

**Figure 4. Correlation between strain and rotation symmetry breaking. a-c,** Strain maps over Figure 3(a) along three reciprocal lattice directions after quadratic background subtraction to remove tip drift. Black dashed lines denote the domain boundary defined in Figure 3(c). **d-f,** Representative -20 mV CDW intensity maps along $q_{2a0}^a$, $q_{2a0}^b$, and $q_{2a0}^c$, respectively. The white arrows in each subfigure denote the corresponding $2a_0$ CDW wavevector. **g,** Smectic vector map over the same region. The smectic order is defined as $n(r) = (I_{2a_0}^a(r) + I_{2a_0}^c(r) - 2I_{2a_0}^b(r), \sqrt{3}(I_{2a_0}^c(r) - I_{2a_0}^a(r)))$, and the angle of each smectic vector is measured with respect to (0,1) direction. **h,** Angle map of (g), with domain wall superimposed. The lower left histogram shows three dominant directions for the smectic vectors. **i,** Map of lattice deformation, with blue lines denoting CDW domain boundary. Isotropic disks represent lattice without strain or with uniform strain, elliptical ones are unidirectionally stretched or compressed according to the main axis of ellipse. White arrows in each domain mark the dominant CDW direction. Pink arrows denote the tensile strain directions locally. **j,** Part of CDW gap map in Figure 3(e), with domain walls marked in black dashed lines. White arrows in each domain serve the same purpose as (i). STM setup condition: $V_{\text{sample}} = 100$ mV, $V_{\text{exc}} = 5$ mV, $I_{\text{set}} = 300$ pA.